\documentclass[a4paper,fleqn,usenatbib]{mnras}
\usepackage[T1]{fontenc}
\usepackage{ae,aecompl}
\usepackage{graphicx}
\usepackage{mathrsfs}
\usepackage{amsmath}
\usepackage{amssymb}
\usepackage{verbatim}
\usepackage{multirow}
\usepackage{subfig}
\usepackage{mathtools}
\usepackage{xspace}

\def\apjl{ApJ}
\def\apjl{ApJ}                
\def\apjs{ApJS}               
\def\ao{Appl.~Opt.}           
\def\aap{A\&A}                

\def \inte {{\em INTEGRAL}}
\def \rxte{{\em RXTE}}

\def \chandra {{\em Chandra}}
\def \xmm{{\em XMM-Newton}}

\def \igr{{IGR J00291$+$5934}}
\def \saxj{{SAX J1808.4$-$3658}}

\def \nustar{{\em NuSTAR}}

\title[Spectral and timing properties of IGR J00291+5934]{Spectral and timing properties of IGR J00291+5934 during its 2015 outburst}
\author[Sanna et al. ]{A. Sanna$^{1}$\thanks{E-mail:
    andrea.sanna@dsf.unica.it},F. Pintore$^{2}$, E. Bozzo $^{3}$, C. Ferrigno$^{3}$, A. Papitto$^{4}$, A. Riggio$^{1}$,\newauthor 
 T. Di Salvo$^{6,5}$, R. Iaria$^{6}$, A. D'A\`i$^{7}$, E. Egron$^{8}$ , L. Burderi$^{1,5}$\\
$^{1}$Dipartimento di Fisica, Universit\`a degli Studi di Cagliari, SP Monserrato-Sestu km 0.7, 09042 Monserrato, Italy\\
$^{2}$INAF-Istituto di Astrofisica Spaziale e Fisica Cosmica - Milano, via E. Bassini 15, I-20133 Milano, Italy\\
$^{3}$ISDC, Department of astronomy, University of Geneva, chemin d'\'Ecogia, 16 CH-1290 Versoix, Switzerland\\
$^{4}$INAF, Osservatorio Astronomico di Roma, Via di Frascati 33, I-00044, Monteporzio Catone (Roma), Italy\\
$^{5}$INFN, Sezione di Cagliari, Cittadella Universitaria 09042, Monserrato, Italy\\
$^{6}$Universit\`a degli Studi di Palermo, Dipartimento di Fisica e Chimica, via Archirafi 36, 90123 Palermo, Italy\\
$^{7}$INAF/IASF Palermo, via Ugo La Malfa 153, I-90146, Palermo, Italy\\
$^{8}$INAF-Osservatorio Astronomico di Cagliari, Via della Scienza 5 - I-09047 Selargius (CA), Italy}
\begin{document}

\date{Accepted -. Received -; in original form -}

\pagerange{\pageref{firstpage}$-$\pageref{lastpage}} \pubyear{2016}

\maketitle

\label{firstpage}

\begin{abstract}
We report on the spectral and timing properties of the accreting millisecond X-ray pulsar \igr{} observed by \xmm{} and \nustar{} during its 2015 outburst. The source is in a {\it hard state} dominated at high energies by a comptonization of soft photons ($\sim0.9$ keV) by an electron population with kT$_e\sim30$ keV, and at lower energies by a blackbody component with kT$\sim0.5$ keV. A moderately broad, neutral Fe emission line and four narrow absorption lines are also found. 
By investigating the pulse phase evolution, we derived the best-fitting orbital solution for the 2015 outburst. Comparing the updated ephemeris with those of the previous outbursts, we set a $3\sigma$ confidence level interval $-6.6\times 10^{-13}$ s/s $< \dot{P}_{orb} < 6.5 \times 10^{-13}$ s/s on the orbital period derivative. Moreover, we investigated the pulse profile dependence on energy finding a peculiar behaviour of the pulse fractional amplitude and lags as a function of energy. We performed a phase-resolved spectroscopy showing that the blackbody component tracks remarkably well the pulse-profile, indicating that this component resides at the neutron star surface (hot-spot).  

\end{abstract}

\begin{keywords}
Keywords: X-rays: binaries; stars:neutron; accretion, accretion disc, \igr{}
\end{keywords}

\section{Introduction}

\igr{} is a transient low mass X-ray binary system (LMXB) observed in outburst for the first time in 2004 \citep[e.g.][]{Eckert2004a, Galloway05a}. After a peculiar double outburst in 2008 \citep[see e.g.][]{Patruno10c,Papitto11b, Hartman2011a}, the source went in outburst again, for the fourth time, in 2015. This last outburst was detected by Swift/BAT on July 23rd 2015 \citep{Sanna2015a}, and lasted approximately 20 days. With its $\sim599$ Hz spin frequency, \igr{} is the fastest objects belonging to the class of systems known as accreting millisecond X-ray pulsars \citep[AMXP; see][for some recent reviews]{Burderi13,Patruno12b}. AMXPs are fast rotating neutron stars (NS) that accrete matter via Roche-lobe overflow from an evolved sub-Solar companion star. The extremely fast spin periods observed in AMXPs are the result of a long-term accretion process in which an old slow-spinning radio-pulsar has continuously gained angular momentum \citep[{\it recycling scenario};][]{Alpar82}. The link between radio millisecond pulsars and AMXPs has been recently confirmed by the radio and X-ray swinging behaviour of AMXPs IGR J18245$-$2452 \citep{Papitto2013b}, as well as for the low luminosity systems PSR J0023+0038 \citep{ Stappers2014a, Archibald2015a} and XSS J12270-4859 \citep{Bassa2014a, Papitto2015a}. 

Timing analysis of the X-ray pulsations revealed a sinusoidal modulation with period of $\sim2.5$ hr and a corresponding projected semi-major axis of $a\, sin(i)\sim65$ lt-ms \citep[see e.g.][]{Galloway05a, Patruno10c, Papitto11b, Hartman2011a}. The NS mass function implies a minimum companion mass of 0.039 M$_\odot$ (assuming a 1.4 M$_\odot$ NS). An upper limit of 0.16 M$_\odot$ for the companion star has been set considering an isotropic a priori distribution of binary system inclinations \citep{Galloway05a}. \citet{Linares07} identified flat-top noise and two quasi-periodic oscillations (QPOs), both at low frequencies (0.01--0.1 Hz), in the source power density spectrum (PDS) obtained by the Rossi X-ray Timing Explorer (\rxte{}) data. While the timing properties of the source have been well studied, the spectral properties are still poorly known. \citet{Falanga05b} found that the source can be described, in the 5-200 keV energy range, by thermal Comptonization with an electron temperature >50 keV. During the 2015 outburst, the source showed for the first time a type-I X-ray burst \citep{Bozzo2015a}, probably ignited in a pure He layer.

Here, we focus on the spectral and temporal properties of \igr{} by analysing high quality, simultaneous \xmm{} and \nustar{} observations of the latest outburst of \igr{}.

\section{Observations and data reduction}
We analysed two simultaneous \xmm{} and \nustar{} observations of the 2015 outburst of \igr{} performed on 2015 July 28 (Obs.ID. 0790181401 and 90101010002, respectively) during the descent phase of the outburst. No type-I bursts were detected.

The EPIC-pn (hereafter PN) and EPIC-MOS2 (hereafter MOS2) cameras were operated in TIMING mode (for a total exposure time of $\sim72$ and $\sim86$ ks, respectively, while the RGS instrument was observed in spectroscopy mode for a total exposure of $\sim86$ ks. We excluded the EPIC-MOS1 (operated in IMAGING mode) from the analysis because highly piled-up. We extracted the \xmm{} data using the Science Analysis Software (SAS) v. 15.0.0 with the up-to-date calibration files, and performing the standard reduction pipeline RDPHA \citep[e.g.][see also XMM-CAL-SRN-0312\footnote{http://xmm2.esac.esa.int/docs/documents/CAL-SRN-0312-1-4.pdf}]{Pintore14}. 
PN and MOS2 source events were selected in the range 0.3$-$10.0 keV and for RAWX=[32:44] and RAWX=[291:322], screening events with \textsc{pattern$\leq$4} and $\leq$12, respectively. Background was extracted in a RAWX region with small source photons contamination. 

We extracted PN and MOS2 energy spectra (source and background) setting \textsc{`FLAG=0'} to retain only events optimally calibrated for spectral analysis. RGS were extracted adopting the \textsc{rgsproc} pipeline, selecting only first order spectra. PN, MOS2 and RGS spectra were binned in order to have at least 100 counts per bin. 

\nustar{} observed \igr{} simultaneously with {\it XMM-Newton}. The observation was reduced performing standard screening and filtering of the events by means of the \nustar{} data analysis software (\textsc{nustardas}) version 1.5.1, for an exposure time of roughly $\sim$40ks for the two instruments. We extracted source and background events from circular regions of 80'' and 120'' radius, respectively. Source spectra, response files and light curves for each instrument were generated using the \textsc{nuproducts} pipeline. 

Solar System barycentre corrections were applied to PN and {\it NuSTAR} photon arrival times with the \textsc{barycen} and {\sc barycorr} tools (using DE-405 solar system ephemeris), respectively. We adopted the source coordinates obtained observing the optical counterpart of the source \citep[][]{Torres08}, and reported in Tab.~\ref{tab:solution}.

\section{Data analysis}

\subsection{Spectral analysis}

\begin{figure}
\centering
\hspace{-0.2cm}
\includegraphics[width=0.48\textwidth]{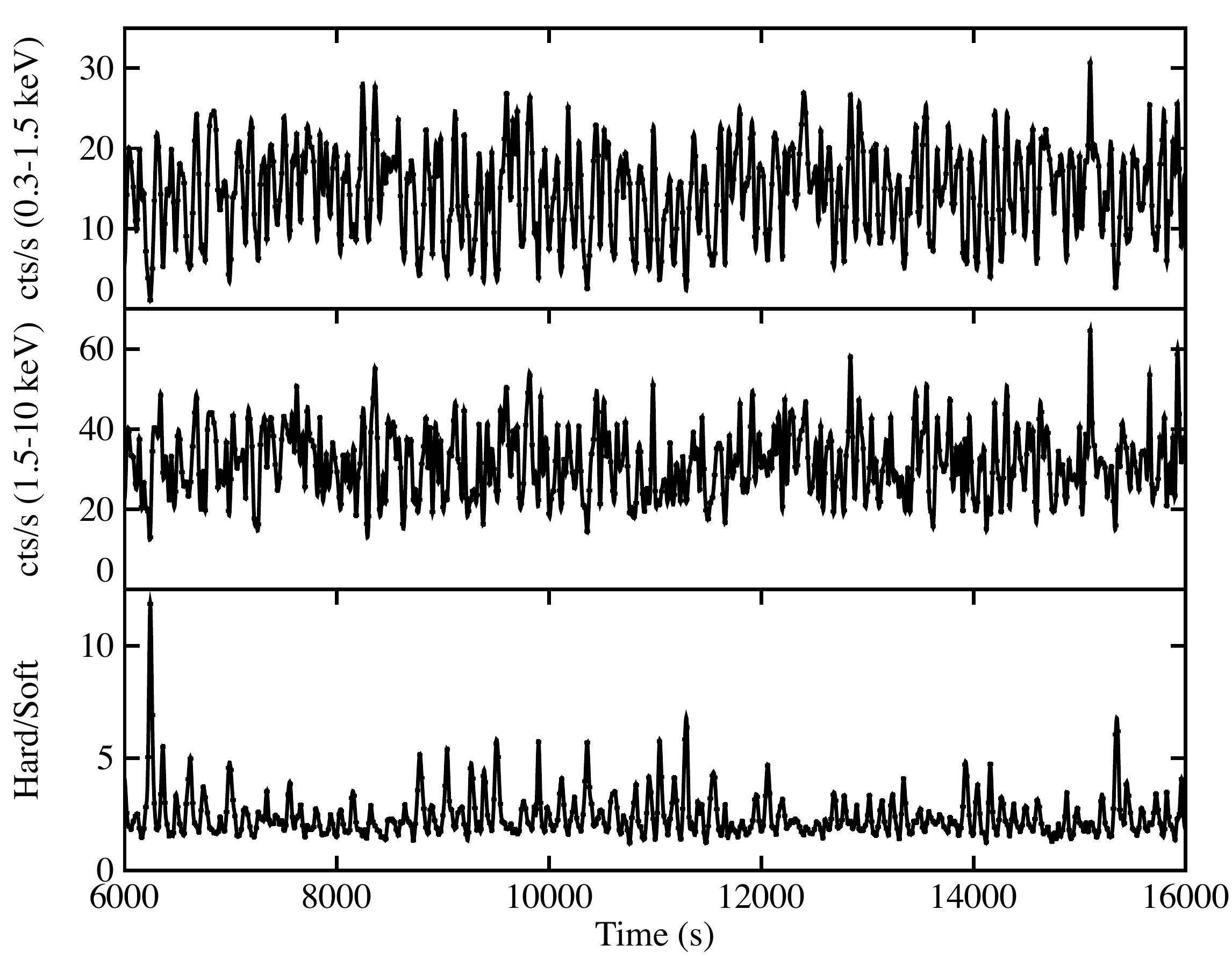}
\caption{Soft (0.3--1.5 keV), hard (1.5--10 keV) and corresponding hardness ratio (hard/soft) PN light-curve of \igr{} .}
\label{fig:lc_hardn}
\end{figure}

The \xmm{} and \nustar{} light-curves of \igr{} shows rapid variability on time-scales between tens and few hundreds seconds, compatible with the low-frequency QPO reported by \citet{Linares07} and \citet{Ferrigno2016a}, and accompanied by spectral variability. Fig.~\ref{fig:lc_hardn} shows a zoom-in of the hardness ratio during the \xmm{} observation.
We suggest the presence of different spectral states as the result of rapid changes in the mechanism generating the hard and the soft emission. Alternatively, the source could be experiencing continuous dipping activity, although quite unlikely as dips usually occur for system seen at high inclination angles \citep[$>70^{\circ}$][]{frank87}. Instead \citet{Torres08} suggested an orbital inclination $22^\circ < i < 32^\circ$, obtained from the analysis of the H-$\alpha$ emission line profile. Moreover, if the frequency of the variability is associated with the Keplerian velocity of a cusp in the outer disc edge, the latter should be at a distance of $\sim0.11$ lt-s which is at least a factor of 1.7 larger than the projected NS semi-major axis assuming an inclination angle $i \leq 65^\circ$ (see Table~\ref{tab:solution}). Although very intriguing, the study of this behaviour is beyond the scope of this paper and it has been instead investigated in a companion paper \citep{Ferrigno2016a}. Here, we focus on the average properties of \igr{}.

We fitted the \xmm{} and \nustar{} average broadband spectrum in the energy range 0.4--10 keV and 3.0--70 keV, respectively, adopting a continuum model based on an absorbed {\sc nthcomp} and a {\sc bbody} component. For the absorber we adopted the {\sc tbabs} model with the abundances of \citet{anders89}. In addition, we modelled the Au instrumental edge ($\sim$2.2 keV) with a {\sc gaussian} component. Interestingly, we observed a broad emission at $\sim6.4$ keV (likely associated to the Fe), fitted with a {\sc gaussian}. 
 
We found the source in a {\it hard} state and we estimated a total unabsorbed source flux in the 0.3--70 keV energy range of $(3.37\pm0.01)\times10^{-10}$ erg cm$^{-2}$ s$^{-1}$. The spectral parameters are characterised by an $N_{\rm H}$ of $(0.226\pm0.007)\times 10^{22}$ cm$^{-2}$ and with the hard energy emission dominated by Comptonization ($\Gamma=1.735\pm0.008$) produced by an electron population with temperature of $27.9_{-3.2}^{+4.8}$ keV. The {\sc bbody} temperature ($0.50\pm0.02$ keV) is not compatible with the temperature of the seed photons of the hard component ($0.85\pm0.05$ keV). Moreover, we note that the PN spectrum shows a different slope above $\sim7$ keV, in comparison with the MOS2 and the \nustar{} spectra. This discrepancy is likely the result of a still uncertain cross-calibration between the PN operated in timing mode and the other instruments. We corrected for this issue by linking the photon index of the \nustar{} and MOS2 spectra and by letting this parameter free to vary with respect to the PN spectrum. As a result we obtained two photon indexes differing by $\sim6-7\%$ (i.e. $1.62\pm0.02$ vs $1.734\pm0.007$). However, the reduced $\chi^2$ of the best-fit is still quite high ($\chi^2/\text{d.o.f.} = 3321.86/2738$). Hence, to take into account the large value of the reduced $\chi^2$, we added a systematic uncertainty of 0.25$\%$ to the spectral data. We note that, hereafter, the uncertainties will be reported at 90\% confidence level for each parameter of interest (see table~\ref{tab:bf}). 

We also observed an Fe emission line at an energy of $6.37\pm0.04$ keV with a $\sigma$ of $80\pm$70 eV and an equivalent width of $\sim$20 eV. To further investigate the origin the feature, we substituted the {\sc gaussian} with a relativistic blurred reflection emission line ({\sc diskline}; \citealt{Fabian89}). 
We fixed the emissivity index and the outer disc radius to -2.7 and $10^5$ R$_g$, respectively, as the fit was insensitive to these parameters.
The best-fit ($\chi^2/\text{d.o.f.} = 2702.79/2737$) gives an inner disc radius poorly constrained ($43^{+910}_{-18}$ R$_g$) and an inclination angle highly unconstrained (Tab.~\ref{tab:bf}). These results confirm the presence of the reflection feature but do not allow us to precisely locate the region of origin in the disc. 
We also applied a self-consistent reflection model {\sc rfxconv} \citep{Kolehmainen11} convolved with the relativistic kernel {\sc rdblur}. 
We obtained a best-fit ($\chi^2/\text{d.o.f.} = 2708.45/2737$) which was statistically worse with respect to the previous model, which gives a very low ionisation parameter log$\xi=1.0\pm0.13$ and a small reflection fraction $0.04^{+0.03}_{-0.01}$. The parameters of the component {\sc rdblur} were highly unconstrained, therefore, we fixed the outer radius and emissivity index to $R_{out}=10^5$ R$_g$ and $\beta=-2.7$, respectively. From the fit we found a poorly constrained source inclination ($36^{+30}_{-16}$ degrees). Moreover, we set an upper limit to the inner radius (R$_{in}>130$ R$_{g}$), which is consistent with the large radius expected by the 0.1 keV width of the line. 

\begin{figure}
\centering
\includegraphics[angle=-90.0, width=0.5\textwidth]{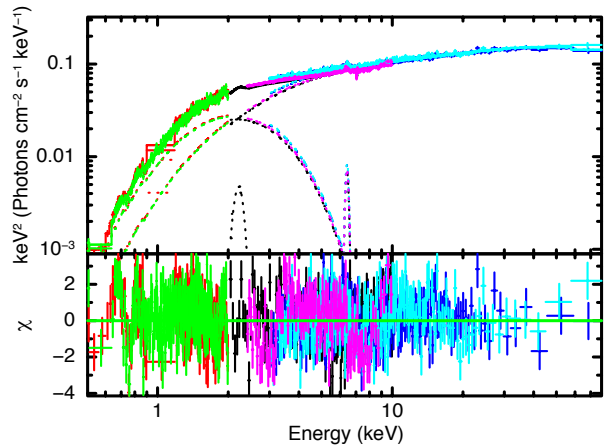}
\caption{\textit{Top-panel:} X-ray spectrum of \igr{} observed by the RGS1 (red points), RGS2 (green points), PN (black points), MOS1 (magenta points), \nustar{} FMPA (blue points), \nustar{} FMPB (cyan points) and the best-fitting model {\sc tbabs*(gaussian+gaussian+bbody+nthcomp)}. \textit{Bottom-panel:} Residuals with respect to the best fitting model. The data have been visually rebinned.}
\label{fig:spectra}
\end{figure}

Finally, we found evidence of some statistically significant absorption lines. In particular, a narrow absorption feature at $7.04\pm0.04$ keV ($\Delta \chi^2=38.1$ for 2 d.o.f., corresponding to F-test probability of chance improvement, $p=1.2\times 10^{-7}$), possibly identified with the Fe XXVI K$_{\alpha}$ transition blue-shifted with v $\sim0.01c$, an absorption lines at $0.756\pm0.002$ keV ($\Delta \chi^2=19.3$ for 2 d.o.f., $p=3.3\times 10^{-4}$) which might be likely associated to O VII or VIII. We also found a marginally significant absorption line at $0.872\pm0.002$ keV which might be associated to Ne I ($\Delta \chi^2=12.23$ for 2 d.o.f., $p=6.2\times 10^{-3}$) and, if realistic, it would be blue-shifted by 0.02-0.03 c. The final best-fit, with these lines and in the case of Fe emission line fitted with a {\sc gaussian} model, gives $\chi^2/\text{d.o.f.} = 2655.67/2730$ and it is shown in Fig.~\ref{fig:spectra}.

\begin{table}

\begin{center}
    \scalebox{0.85}{\begin{minipage}{28.0cm}
\begin{tabular}{llll}
\hline
Model & Component & (1) & (2) \\
\hline
{\sc TBabs} & nH ($10^{22}$) & $0.237^{+0.01}_{-0.009}$ &$0.237^{+0.008}_{-0.007}$ \\
\\
{\sc bbody} & kT (keV) & $0.5^{+0.02}_{-0.02}$ & \\
 & norm (10$^{-4}$) & $6.8^{+0.5}_{-0.6}$ & \\
\\
{\sc nthComp} & Gamma (pn) & $1.62^{+0.02}_{-0.02}$ & \\
 & Gamma (MOS, NuSTAR) & $1.734^{+0.007}_{-0.006}$ & \\
 & kT$_{e}$ (keV) & $27.9^{+4.8}_{-3.2}$ & \\
 & kT$_{bb}$ (keV) & $0.85^{+0.05}_{-0.06}$ &\\
 & norm (10$^{-3}$) & $6.5^{+0.9}_{-0.7}$  & \\
\\
{\sc gaussian} & LineE (keV) & $6.37^{+0.04}_{-0.04}$ & - \\
 & $\sigma$ (keV) & $0.08^{+0.06}_{-0.08}$ & -\\
 & norm (10$^{-5}$) & $4.3^{+1.3}_{-1.2}$ & \\
 \\
{\sc Diskline} & LineE (keV) & - & $6.46^{+0.08}_{-0.06}$ \\
 & Betor10 & - & -2.7 (frozen)   \\
 & Rin(M) & - & $43^{+910}_{-18}$ \\
 & Rout(M)(10$^{5}$) & & 1.0 (frozen)  \\
 & Incl(deg) & - & unconstrained  \\
 & norm(10$^{-5}$) & - & $4.3^{+1.2}_{-1.2}$ \\
\\
\hline
 $\chi^{2}/\text{d.o.f.}$ & & 2702.72/2738 & 2702.79/2737\\
\hline
\end{tabular}
\end{minipage}}
\caption{Best fit spectral parameters obtained with the absorbed continuum {\sc bbody+nthcomp} model, plus either a {\sc gaussian} (model 1) or {\sc diskline} (model 2) component for the Fe emission line. Uncertainties are at 90$\%$ for each parameter of interest.}
\label{tab:bf}
\end{center}
\end{table}

\subsection{Timing analysis}
\label{sec:ta2016}

Following the procedure described in \citet{Burderi07}, we corrected the photon time of arrivals of both the PN and the \nustar{} datasets for the delays caused by the binary motion applying the orbital parameters reported by \citet[][see also \citealt{Patruno10c,Hartman2011a}]{Papitto11b} for the latest outburst of the source on September 2008. We then performed an epoch-folding search of the whole PN and \nustar{} observations around the spin frequency of the September 2008 outburst \citep[][]{Papitto11b}. We found evidence of X-ray pulsation in both the PN and \nustar{} observations. The PN average pulse profile is well fitted by two sinusoids with fractional amplitude of 13.45(7)\% and 0.58(7)\%, for the fundamental and the first harmonic, respectively. The \nustar{} pulse profile is well described by a single sinusoid with fractional amplitude of 11.2(2)\%.

\begin{figure}
\centering
\includegraphics[width=0.45\textwidth]{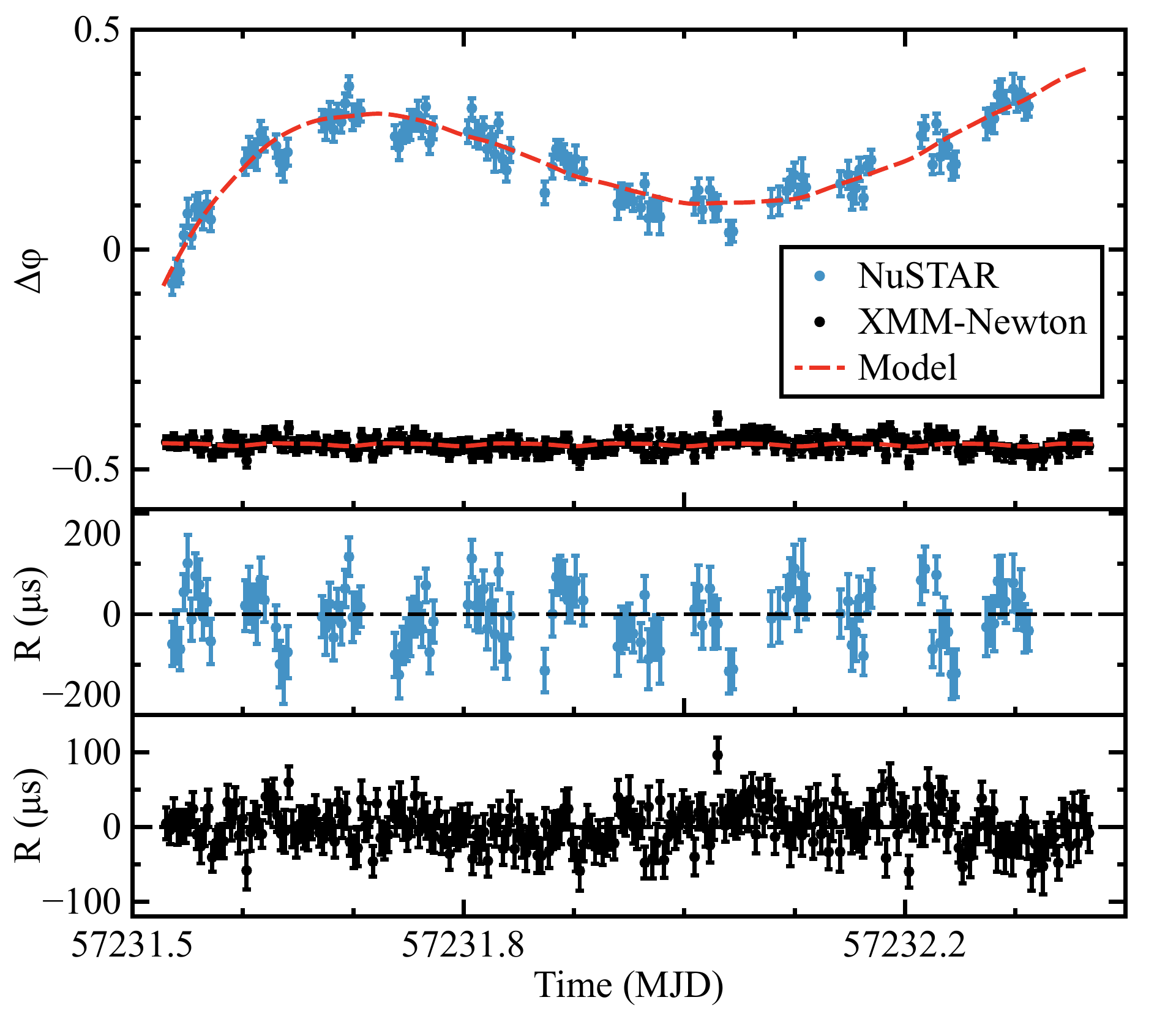}
\caption{\textit{Top panel -} Pulse phase delays as a function of time computed by epoch-folding 300 second data intervals of the \xmm{} observations and 500 second data intervals of the \nustar{} observation. Red dotted line represents the best-fit model (see text for more details). \textit{Middle panel -} Residuals in $\mu s$ of the \nustar{} data with respect to the best-fitting orbital solution. \textit{Bottom panel -} Residuals in $\mu s$ of the \xmm{} data with respect to the best-fitting orbital solution.}
\label{fig:phase_fit}
\end{figure}

We created pulse phase delays computed over time intervals of approximately 300s for the PN observation and 500s for the \nustar{} observation, epoch-folding the intervals at the mean spin frequency values for each instrument. Although strictly simultaneous, no phase-connected timing analysis was applicable due to the presence of a time drift on the internal clock on \nustar{} \citep{Madsen15}, which affects the observed coherent signal making the spin frequency value significantly different compared to the \xmm{} value. It can be shown that the latter spurious phase delay only marginally affects the phase delays induced by the orbital motion \citep[see, e.g.][for similar phenomena]{Riggio07}, therefore we were still able to investigate the orbital ephemerides combining the two observations. 
We fitted the pulse phase delays time evolution with the following models:
\begin{eqnarray}
\label{eq:ph}
\begin{cases}
\Delta \phi_{NuStar}(t)=\sum\limits_{n=0}^{4} \frac{C_n}{n!}(t-T_0)^n+R_{orb}(t)\\
\Delta \phi_{XMM}(t)=\sum\limits_{n=0}^{2} \frac{D_n}{n!}(t-T_0)^n+R_{orb}(t)
\end{cases}
\end{eqnarray}
where the first element represents a polynomial function used to model the phase variations in each dataset. Additionally, the component $R_{orb}(t)$ represents the Roemer delay component, where the differential correction on the orbital parameters are determined combining the two dataset. If a new set of orbital parameters is found, we repeated the process described above until no significant differential corrections were found. We reported the best-fit parameters in Table~\ref{tab:solution}, while in Fig.~\ref{fig:phase_fit} we showed the pulse phase delays of the two observations with the best-fitting models (top panel), and the residuals with respect to the best-fitting model.

We estimated the systematic uncertainty induced on the spin frequency correction $\Delta \nu_0$ because of the positional uncertainties of the source using the approximated expression $\sigma_{\nu_{pos}}\leq \nu_0\,y\,\sigma_{\gamma}(1+\sin^2\beta)^{1/2}2\pi/P_{\oplus}$, where $y=r_E/c$ is the semi-major axis of the orbit of the Earth in light-seconds, $P_{\oplus}$ is the Earth orbital period, and $\sigma_{\gamma}$ is the positional error circle of the residuals \citep[for more details see e.g.][]{Lyne90, Burderi07}. Adopting the positional uncertainties reported by \citet{Torres08}, we estimate $\sigma_{\nu_{pos}} \leq 5\times 10^{-8}$~Hz. We added in quadrature the systematic uncertainty to the statistical errors of $\nu_0$ reported in Table~\ref{tab:solution}.

\begin{figure}
\includegraphics[width=0.45\textwidth]{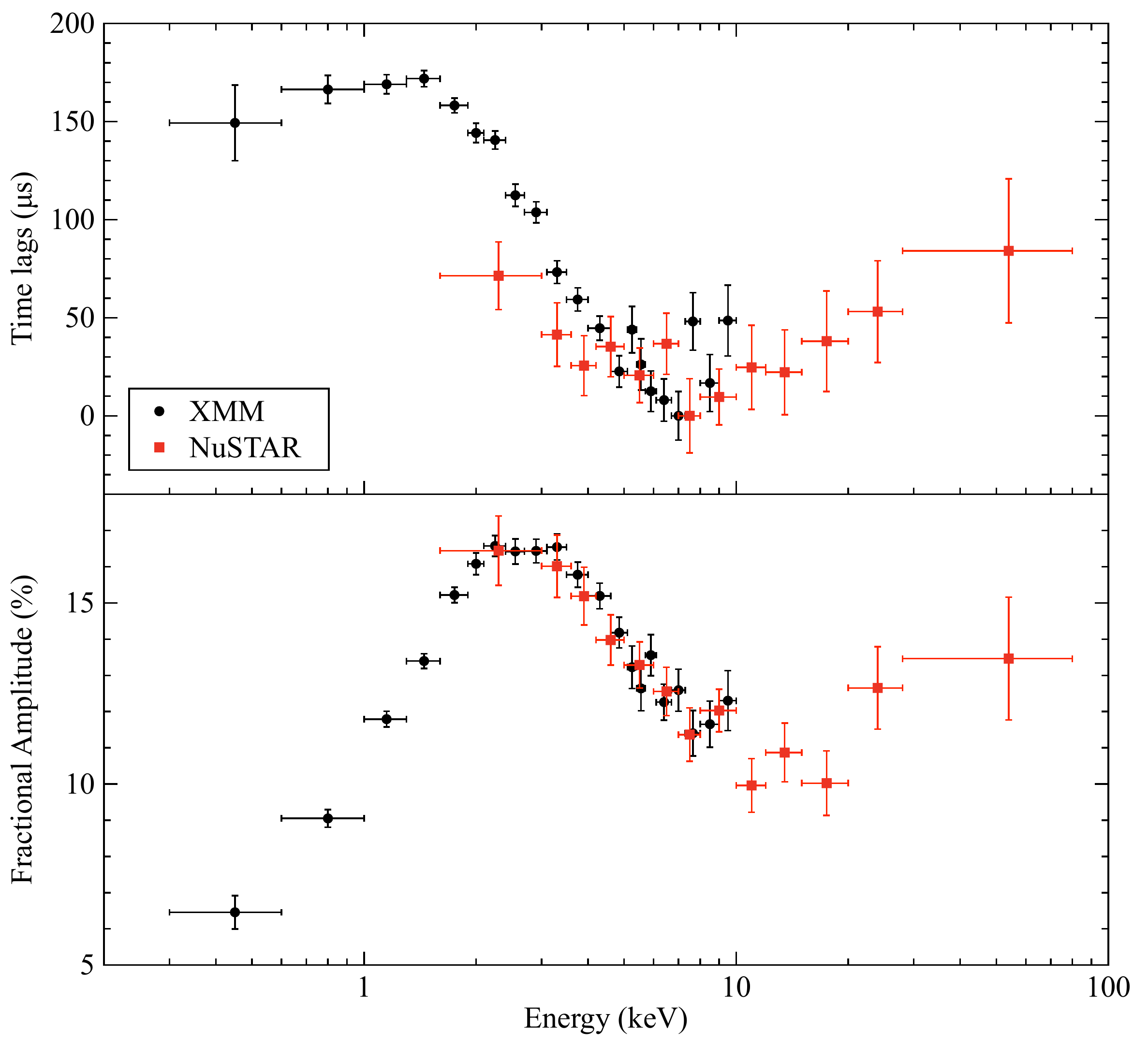}
\caption{\textit{Top panel -} Time lags in $\mu s$ evolution as a function of energy obtained from the \xmm{} and \nustar{} observations reported with black dots and red squares, respectively. For display purpose only the reference pulse profiles of the two datasets has been set around 7 keV. \textit{Bottom panel -} Evolution of the pulse fractional amplitude for the two datasets. The colour coding has been kept fixed between the panels.}
\label{fig:frac_amp}
\end{figure}

Finally, we explored the dependence of the pulse profile fractional amplitude and time lags (defined as the phase shift of the fundamental harmonic of the pulse profile in different energy bands) as a function of energy, dividing the energy range between 0.3 keV to 10 keV into 21 intervals, and the energy range between 1.6 keV and 80 keV into 13 intervals, for \xmm{} and \nustar{}, respectively. Since the two dataset are not phase connected, we calculated the time lags using different reference profiles. For display purposes only, for both dataset we set the reference at around 7 keV. In Fig.~\ref{fig:frac_amp} we reported the dependence of the time lags (top panel) and the fractional amplitude (bottom panel) as a function of energy for the \xmm{} and \nustar{} pulse profiles, represented with black dots and red squares, respectively.

\begin{table}
\begin{center}
\begin{tabular}{l | c  c}
Parameters             & XMM & \nustar{} \\
\hline
\hline
R.A. (J2000) &  \multicolumn{2}{c}{$00^h29^m3.05^s$}\\
DEC (J2000) & \multicolumn{2}{c}{$59^\circ34^m18.93^s$}\\
$P_{orb}$ (s) & \multicolumn{2}{c}{8844.08(2)}\\
$x$ (lt-s) &\multicolumn{2}{c}{0.0649905(24)} \\
$T_{NOD}$ (MJD) & \multicolumn{2}{c}{57231.437581(3)}\\
e &\multicolumn{2}{c}{<$2\times 10^{-4}$}\\
$\nu_0$ (Hz) &598.8921309(2)&598.892168(2) \\
$\dot{\nu}_0$ (Hz/s) &$3(5)\times 10^{-12}$&- \\
$T_{ref}$ (MJD) & \multicolumn{2}{c}{57231.5$^{a}$}\\
\hline
$\chi^2_\nu$/\text{d.o.f.} & \multicolumn{2}{c}{491.5/348	}\\
\end{tabular}
\caption{Orbital parameters and spin frequency of \igr{} obtained from the analysis of the \xmm{} and \nustar{} observations of the source. Errors are at 1$\sigma$ confidence level. Uncertainties are also scaled by a factor $\sqrt{\chi^2_{red}}$ to take into account the large value of the reduced $\chi^2$. The reported X-ray position of the source has a pointing uncertainty of 0.05$''$ \citep[see e.g.][]{Torres08}.$^a$ Timing solution reference epoch.}
\label{tab:solution}
\end{center}
\end{table}

\section{Phase-resolved spectroscopy}

\begin{figure}
\centering
\includegraphics[width=0.45\textwidth]{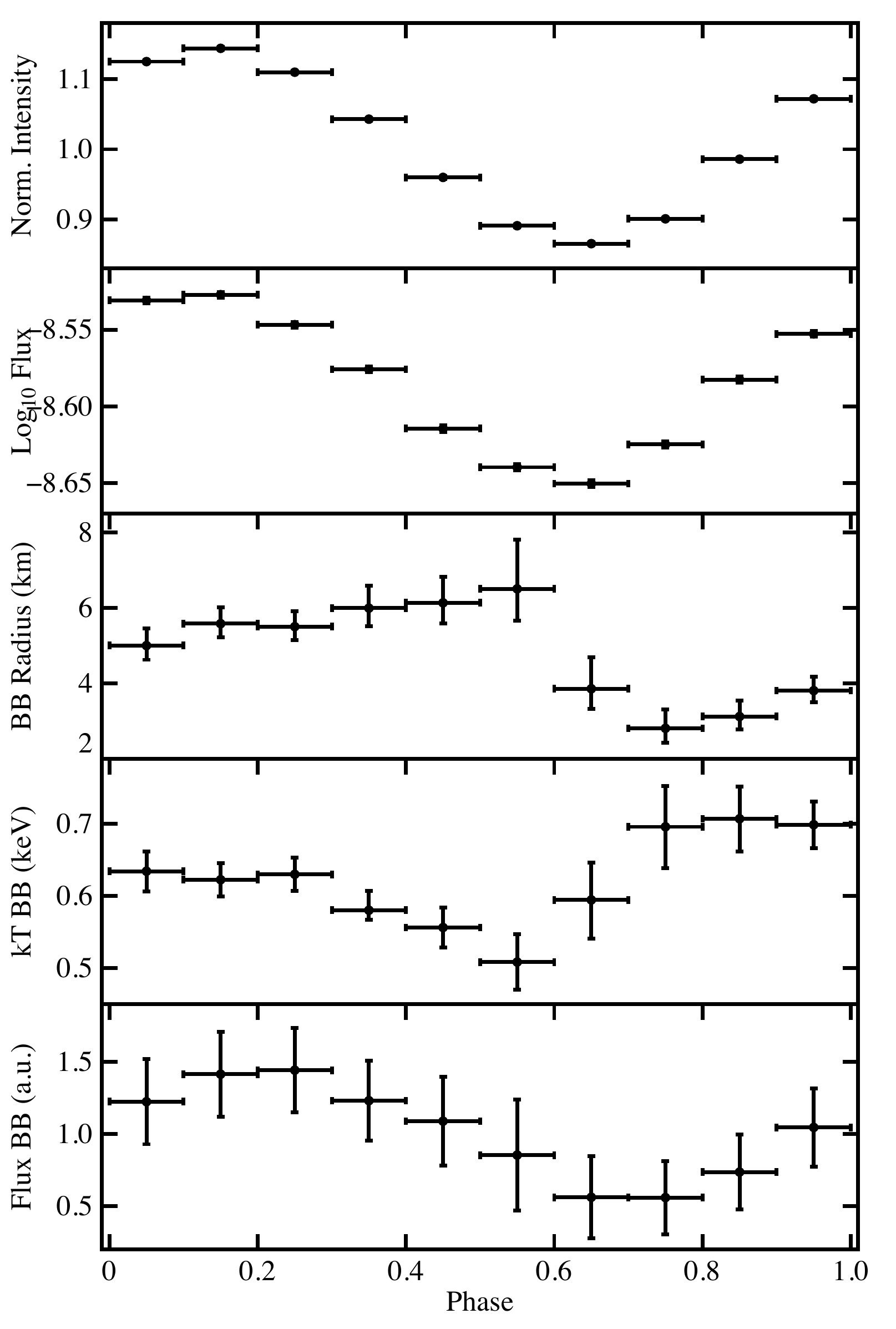}
\caption{Phase-resolved spectral analysis as reported in Table~\ref{tab:pr2}. In the top panel, the pulse profile in the 0.3--10 keV energy band, in the second panel the unabsorbed 0.3--10 keV flux (in erg cm$^2$ s$^{-1}$), in the third and fourth panels the {\sc bbodyrad} temperature and the corresponding radius, and in the bottom panel the blackbody flux estimated as $r^2kT^4$ and expressed in arbitrary units (a.u.). Uncertainties are expressed at a 90\% confidence level.}
\label{pr1}
\end{figure}

After correcting the photon times of arrival for the orbital ephemerides reported in Table~\ref{tab:solution}, we performed a phase-resolved spectroscopy with the aim to study the evolution of the spectral components as a function of the NS spin phase. Given the previously described spectral discrepancy between the detectors and the timing issues reported for \nustar{}, we decided to focus on the PN data only.
We split the pulse profile in 10 phase bins of equal size and for each phase interval we extracted the corresponding PN spectrum. We fitted simultaneously the 10 spectra in the energy range 2.0--10 keV. In analogy with the best-fit model of the average broadband spectrum, we fitted the continuum with the model {\sc bbodyrad+nthcomp}. Moreover, we included two broad emission features to take into account the Au instrumental feature at 2.2 keV and the Iron line. Given the lack of coverage below 2 keV, we fixed the hydrogen column density to the value $0.237\times10^{22}$ cm$^{-2}$ and the kT$_e$ temperature of the comptonisation component ({\sc nthcomp}) to 27.8 keV (i.e., the best-fit values of the averaged spectrum). 
We note that the high energy component was only marginally changing among the spectra, therefore we decided to link together the parameters of the 10 spectra. 
On the other hand, the blackbody component clearly showed variability correlated with the spin pulse profile (see Table~\ref{tab:pr2} and Fig.~\ref{pr1}), with variations of 0.2 keV. In particular we found that the temperature profile is slightly offset with respect to the pulse-profile, with the highest/lowest blackbody temperature lagging by $\sim$0.1 in phase the peak/bottom of the sinusoidal pulse-profile. A similar result is also found for the corresponding emitting radius in the opposite direction, so that the black-body bolometric flux remain aligned with the pulse profile, with the large uncertainties. These results allow us to unequivocally associate the soft component to the NS.

\begin{table*}
\begin{center}
    \scalebox{0.8}{\begin{minipage}{28.0cm}
\begin{tabular}{llllllllllll}
\hline
Model & Component & (1) & (2) & (3) & (4) & (5) & (6) & (7) & (8) & (9) & (10) \\
\hline
{\sc TBabs} & nH ($10^{22}$) &  \multicolumn{10}{c}{0.226 (fixed)} \\
\\
{\sc bbodyrad} & kT (keV) & $0.63_{-0.03}^{+0.03}$ & $0.62_{-0.02}^{+0.02}$ & $0.63_{-0.02}^{+0.02}$ & $0.58_{-0.01}^{+0.03}$ & $0.56_{-0.03}^{+0.03}$ & $0.51_{-0.04}^{+0.04}$ & $0.59_{-0.05}^{+0.05}$ & $0.70_{-0.06}^{+0.06}$ & $0.71_{-0.05}^{+0.04}$ & $0.70_{-0.03}^{+0.03}$ \\
 & norm & $156.6_{-23.5}^{+28.2}$ & $195.0_{-25.7}^{+30.1}$ & $189.5_{-24.3}^{+28.4}$ & $224.9_{-35.5}^{+44.3}$ & $235.5_{-41.5}^{+53.2}$ & $264.8_{-69.1}^{+106.5}$ & $93.2_{-26.1}^{+40.3}$ & $49.5_{-13.2}^{+17.8}$ & $61.1_{-13.4}^{+16.6}$ & $90.8_{-15.0}^{+17.6}$ \\
\\
{\sc nthComp} & $\Gamma$ & \multicolumn{10}{c}{$1.459^{+0.02}_{-0.009}$} \\
 & kT$_e$ (keV) & \multicolumn{10}{c}{28.8 (fixed)} \\
 & kT$_{bb}$ (keV) & \multicolumn{10}{c}{$0.09^{+0.18}_{-0.09}$} \\
 & norm & $0.29_{-0.06}^{+0.01}$ & $0.29_{-0.06}^{+0.01}$ & $0.28_{-0.06}^{+0.01}$ & $0.26_{-0.05}^{+0.01}$ & $0.24_{-0.05}^{+0.005}$ & $0.232_{-0.05}^{+0.004}$ & $0.233_{-0.05}^{+0.009}$ & $0.25_{-0.05}^{+0.01}$ & $0.27_{-0.06}^{+0.01}$ & $0.28_{-0.06}^{+0.01}$ \\
\\
 {\sc gaussian} & LineE(keV) &  \multicolumn{10}{c}{$6.47^{+0.04}_{-0.04}$} \\
 & Sigma(keV) & \multicolumn{10}{c}{$0.13^{+0.04}_{-0.04}$} \\
 & norm(10$^{-4}$) & \multicolumn{10}{c}{$5.1^{+1.0}_{-1.0}$} \\
 \\
 {\sc gaussian} & LineE(keV) &  \multicolumn{10}{c}{$2.240^{+0.01}_{-0.006}$} \\
 & Sigma(keV) & \multicolumn{10}{c}{$0.04^{+0.02}_{-0.03}$} \\
 & norm(10$^{-3}$) & \multicolumn{10}{c}{$1.2^{+0.3}_{-0.3}$} \\
\hline
&  $\chi^{2}/\text{d.o.f.}$ &  \multicolumn{10}{c}{976.42/957}\\
\hline
\end{tabular}
\end{minipage}}
\caption{Best fit spectral parameters obtained with an absorbed continuum {\sc bbody+nthcomp} model, plus either two {\sc gaussian} components for the Fe emission line and the instrumental Au calibration residual. Uncertainties are expressed at a 90\% confidence level}
\label{tab:pr2}
\end{center}
\end{table*}

\section{Discussion}

\igr{} shows a hard spectrum dominated by comptonization as typically  observed in other AMXPs \citep[e.g.][]{Papitto09,Patruno10a}.
The region producing the seed photons can be estimated by using the relations proposed by \citet[][$R_{seed}=3\times10^4\,d\sqrt{f_{bol}/(1+y)}/kT_{bb}^2 $]{in-t-Zand1999a}, where $f_{bol}$ is the bolometric flux of the {\sc nthcomp} component and {\it y} is the Compton parameter. 
Assuming a distance of 4 kpc \citep{Galloway05a}, we inferred a region size of $1.4\,d_{4kpc}$ km.
The soft component shows instead a temperature of 0.5 keV, not compatible with the seed photons temperature, and corresponding to an emission region of $0.5\,d_{4kpc}$ km. The size of this region is almost a factor of three smaller compared with the seed photons, however, both are likely incompatible with the inner radius of the accretion disc. Our interpretations are quite consistent with the findings of \citet{Paizis05} which performed a spectral analysis on \chandra{} and \rxte{} data of the fainter end of the 2005 source outburst.

Remarkably, the phase-resolved spectroscopy analysis allowed us to clearly associate the blackbody component with the NS, as we found that the variations in temperature are related to the pulse evolution. We associated the blackbody component with the hot-spot on the NS surface, and its radius and temperature track well the evolution of the pulse profile, indicating that the temperature is higher or lower when the pulse points toward us or in opposition to us. 
We note that the delay of $\sim$0.1 in phase of the temperature with respect to the pulse-profile may be caused by comptonization effects of the hot-spot thermal emission. Alternatively, a combination of temperature gradient in the hot-spot emission region and inclination angle of the accretion column with respect to the line of sight could be responsible for the observed phase misalignment. More observations with higher statistics are required to further investigate the spectral properties of the source as function of the spin pulse phase in order to confirm or disprove the proposed scenario.

The presence of a neutral Fe emission line only weakly broadened (0.1 keV, i.e. $\sim$0.01c assuming Doppler broadening) combined with the lack of statistically significant improvement when applying a self-consistent relativistically smeared reflection component to the data do not robustly support a reflection originated from the inner regions of the disc. Nonetheless, we cannot discard the possibility of a line originated by reflection off of hard photons in the outer disc. However, since persistent X-ray pulsations require a magnetospheric radius smaller than the co-rotation radius (of the order of 24 km for \igr{}), it needs to be explained the lack of the reflection component in this region. 
We suggest two possible scenarios to explain such findings: i) if the line is produced by reflection off of hard photons by the accretion disc, the self-consistent model suggest a weakly ionised disc, consistent with the presence of neutral iron, but produced very far from the NS ($>100$ R$_g$) and this may be compatible with the phenomenology of systems accreting at low Eddington rates \citep[$<10^{-2}$ L$_\text{edd}$, see][for more details on the topic; see also \citealt{DAi10}, \citealt{Di-Salvo2015a}]{Degenaar2016a}. We suggest that the lack of inner disc reflection may be due to either a continuum inefficient to illuminate the disc as a beamed emission of the direct comptonisation component inclined with respect to the accretion disk, or the inner disc may eject outflows (in the form of winds) for local violations of the Eddington limit \citep[e.g.][]{Poutanen2007a}, that might also be responsible for the absorption lines observed in the system. 
This could cause changes in the amount of mass transferred onto the NS giving raise to the observed variability on timescales of $\sim100$s, in analogy with the Galactic BH candidate GRS 1915+105 and to its observed ``heartbeat'' on short-timescales and also outflowing wind \citep[e.g.][]{Neilsen2012a}. 
However, a caveat on this interpretation may be the luminosity of \igr{}, which is orders of magnitude lower than that of GRS 1915+105, hence making difficult to explain how such outflows are produced. However, similar "heartbeats" are also observed in the BH candidate IGR J17091-3624 \citep{Altamirano2011b} which may have a luminosity orders of magnitude lower than that of GRS 1915+105, hence allowing not to exclude such processes also in IGR J00291+5934.
ii) Alternatively, the iron line may be produced in a corona and broadened
either by velocity dispersions (of the order of 0.01 c) in the corona or
by Compton scattering in a moderately optically thick and relatively cold
medium; in the latter case also the mechanism proposed by \cite{Laurent2007a} of a Compton down-scattering produced by a narrow wind shell outflowing at mildly relativistic speed may play a role, although in this case a significant red-wing of the line is also expected.

\begin{figure}
\centering
\includegraphics[width=0.45\textwidth]{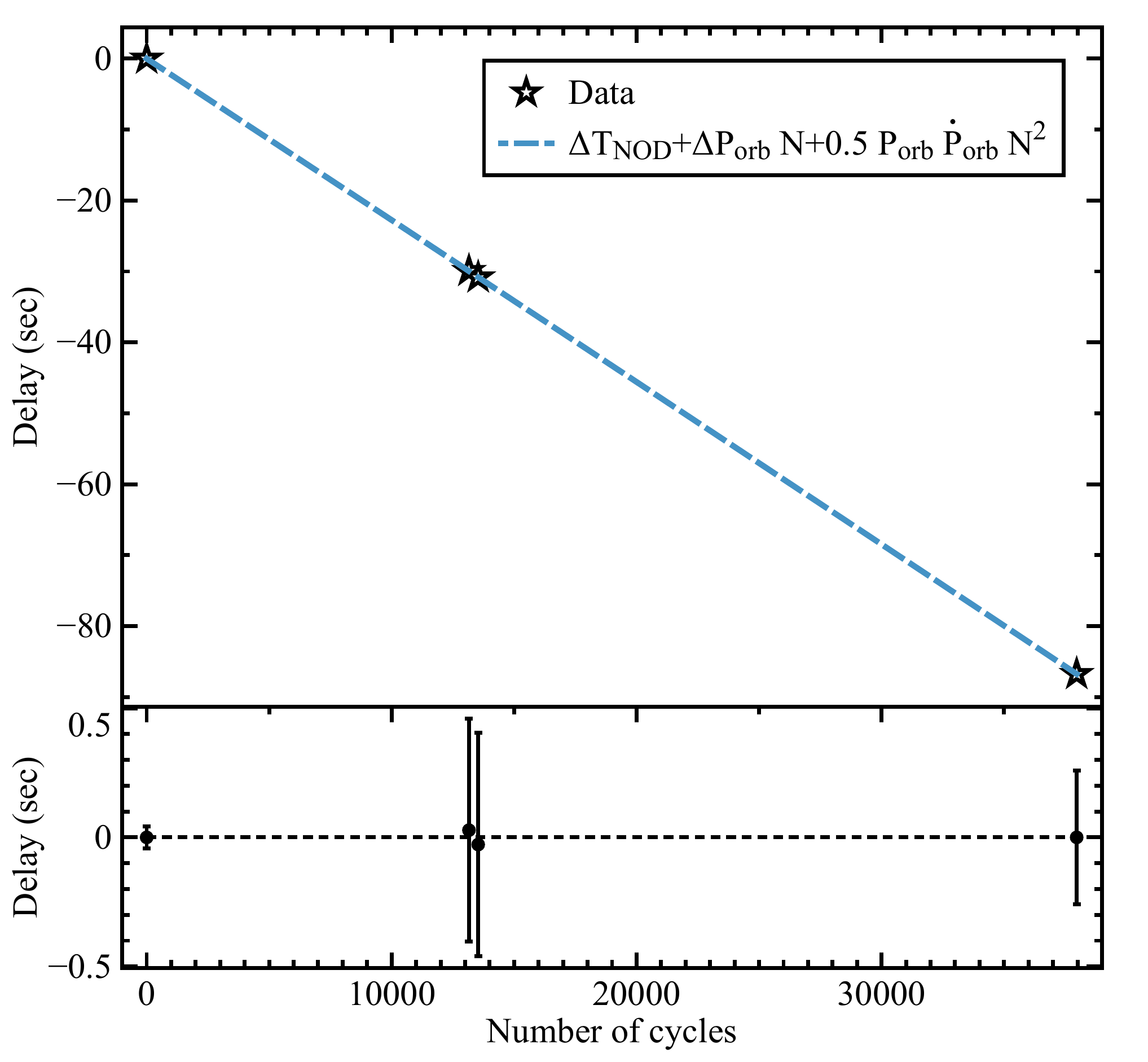}
\caption{\textit{Top panel -} Time delays of the NS time of passage from the ascending node for each of the observed outbursts of \igr{}. The cyan dashed line represents the best-fitting parabola used to model the data. \textit{Bottom panel -} Residuals in seconds of the time delays with respect to the best-fitting timing solution.}
\label{fig:fit_tstar}
\end{figure}

From the timing analysis, we detected X-ray pulsations at the spin frequency period of \igr{} with an average fractional amplitude of 13\% and 11\% for \xmm{} and \nustar{}, respectively. From the evolution of the pulse phase delays obtained from the two datasets of the 2015 outburst, we obtained an updated timing solution for the source. The new set of orbital parameters is compatible within the errors with the previous timing solution obtained from the analysis of the 2004 \citep{Galloway05a, Falanga05b, Burderi07} and 2008 outbursts \citep{Patruno10c, Papitto11b, Hartman2011a}.
Combining the measurements of the time of passage of the NS at the ascending node ($T_{NOD}$) for each observed outburst we computed the delays $\Delta T_{NOD}$ as a function of time. $\Delta T_{NOD}$ is obtained by subtracting from each measurement the $T_{NOD_{pred}}=T_{NOD_{ref}}+N P_{orb_{ref}}$ predicted by a constant orbital period model \citep[see e.g.][]{diSalvo08, Burderi09, Sanna2016a} adopting the orbital period $P_{orb_{ref}}=8844.079(1)$ s and the time of passage of the NS at the ascending node $T_{NOD_{ref}}=53345.1619264(5)$ MJD of the first observed outburst of the source in 2004 reported by \citet[][]{Papitto11b}. The integer $N$
represents the closest integer number of orbital cycles elapsed between two different $T_{NOD}$ observed. We fitted the delays with the expression $\Delta T_{NOD} = \delta T_{NOD_{ref}} + N\, \delta P_{orb_{ref}}+0.5\,N^2\, \dot{P}_{orb}P_{orb_{ref}}$,
where $\dot{P}_{orb}$ represents the orbital period derivative. We obtained the best-fit for $\delta T_{NOD_{ref}}= -0.0005(4) \times 10^{-2}$ MJD, $\delta P_{orb_{ref}} = 2.27(4)\times 10^{-3}$ s and $\dot{P}_{orb}=-0.7(2.2)\times 10^{-13}$ s/s, with $\chi^2=0.008$ for 1 degree of freedom. Fig.~\ref{fig:fit_tstar} shows $\Delta T_{NOD}$ values for the observed outbursts as a function of the number of orbital cycles elapsed from the reference epoch (top panel) as well as the residuals with respect to the best-fitting model described above (bottom panel). We note that the reduced $\chi^2$ is significantly smaller compared with the expected value one, likely indicating that the testing of the model would require more precise data. However, the probability to obtain a $\chi^2$ smaller than the observed is $\sim7\%$, marginally above the conventionally accepted significant threshold of 5$\%$ \citep[see e.g.][]{Bevington2003a}.
From the fit, we cannot constrain both the correction on the time of passage from the ascending node and the orbital period derivative. However, we find a significant correction on the orbital period at the reference time (2004 outburst) finding $P_{orb}=8844.07673(3)$ s, in agreement with the estimate reported by \citet{Hartman2011a}. Moreover, we note that the timing solution and our new estimate of the orbital period found from our analysis are fully consistent, within uncertainties, with those obtained from the analysis of the INTEGRAL data \citep{DeFalco2016a}. For the orbital period we can at least define the $3\sigma$ confidence level interval $-6.6\times 10^{-13}$ s/s $< \dot{P}_{orb} < 6.5 \times 10^{-13}$ s/s \citep[compatible within errors with the estimate recently reported by][]{Patruno2016b}. 

\begin{figure}
\centering
\includegraphics[width=0.45\textwidth]{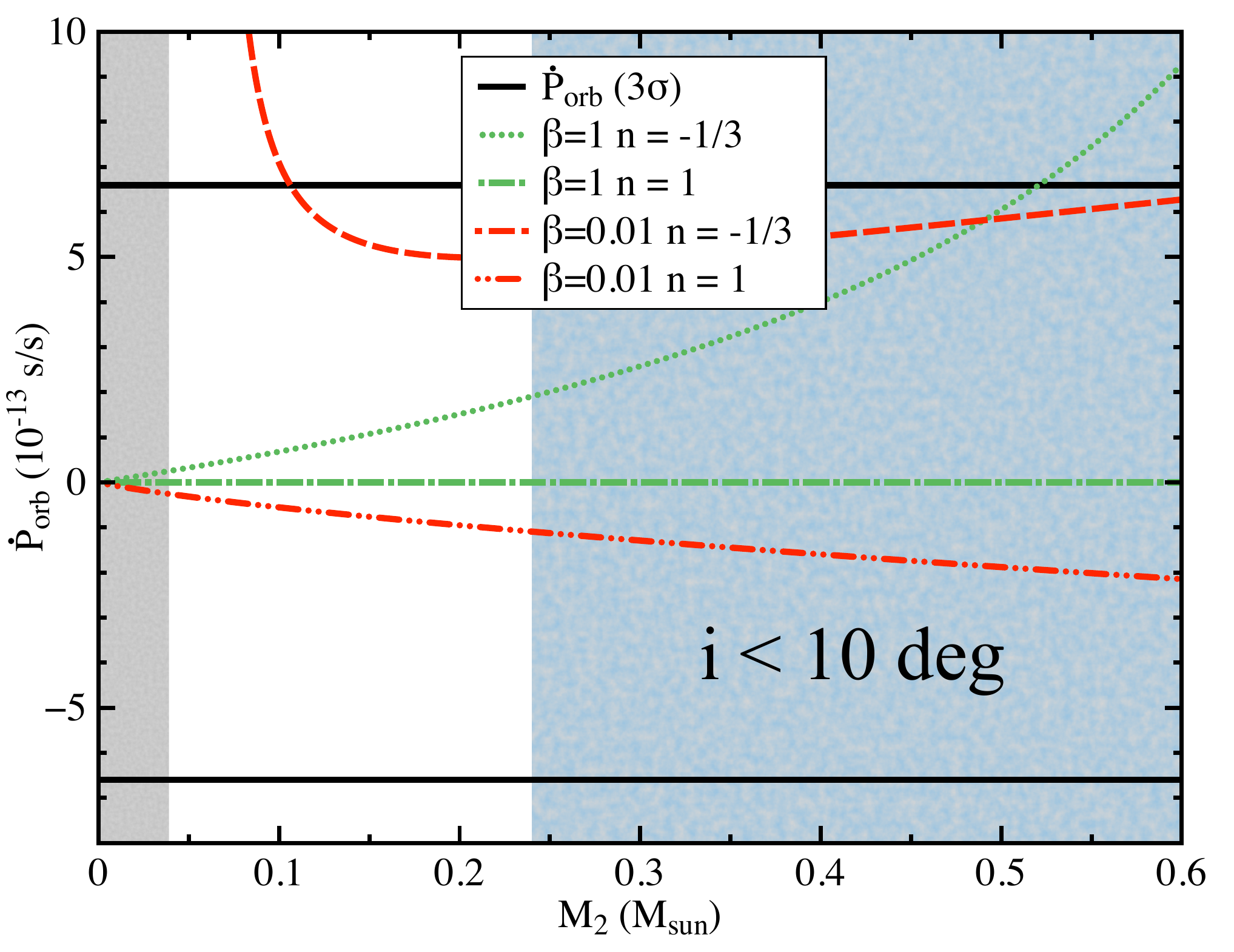}
\caption{Orbital period derivative versus companion star mass in the hypothesis of conservative and non-conservative mass transfer (with mass leaving the system with the specific angular momentum at the inner Lagrangian point). Dotted and dashed-dotted green curves represent the conservative ($\beta=1$) mass-transfer scenario assuming a low-main sequence ($n=1$) and a fully convective ($n=-1/3$) companion star, respectively. Dashed and dot-dot-dashed red curves represent the non-conservative ($\beta= t_{out}/t_{quiet}\simeq 0.01$) mass-transfer scenario again for a low-main sequence ($n=1$) and a fully convective ($n=-1/3$) companion star, respectively. The horizontal black lines delimit the $3\sigma$ orbital period derivative confidence level interval obtained from the analysis. Finally, vertical lines represent the minimum mass estimated from the binary mass function (for a 1.4  M$_{\odot}$ NS) and the companion mass for an almost face-on system (i=$10^\circ$).}
\label{fig:orb_ev}
\end{figure}

Spin-orbit coupling scenarios, such as the model proposed by \citet{Applegate1992a} and \citet{Applegate1994a} to explain the peculiar secular orbital evolution of a small group of black widow millisecond pulsars \citep[see e.g.,][]{Arzoumanian1994a, Doroshenko2001a} have also been invoked to described the orbital evolution of the AMXP \saxj{} \citep{Hartman08, Hartman09b, Patruno12a}. However, as noted also by \citet{Patruno2016b}, the very similar orbital parameters of \igr{} show a complete different secular orbital evolution when compared with \saxj{}. Here we will discuss the orbital evolution of \igr{} in the light of mechanism proposed by \citet{diSalvo08} and \citet{Burderi09} for the AMXP \saxj{}.
Following \citet{diSalvo08}, we can describe the orbital period derivative caused by mass transfer induced by emission of gravitational waves as
\begin{equation}
\label{eq:evo}
\dot{P}_{orb}=-1.38\times10^{-12}m_1^{5/3}q(1+q)^{-1/3}P_{2h}^{-5/3}\Big[\frac{n-1/3}{n-1/3+2g}\Big]s/s,
\end{equation}
where $m_1$ is the NS mass in units of M$_{\odot}$, $q=m_2/m_1$ is the binary mass ratio, $m_2$ is the companion mass in units of M$_{\odot}$, $P_{2h}$ is the binary orbital period in units of two hours, $n$ is the mass-radius index, $g=1-\beta q-(1-\beta)(\alpha+q/3)/(1+q)$ and $\alpha$ is the specific angular momentum of the mass ejected in units of the specific angular momentum of the secondary.
In Fig.~\ref{fig:orb_ev}, we show equation~\ref{eq:evo} for different values of $\beta$ and mass index $n$. In particular, we investigate a conservative ($\beta=1$) scenario and a non-conservative scenario in which the mass-loss rate from the secondary is kept constant. In the latter scenario we assume that mass lost from the companion is accreted during outbursts and ejected (from the internal Lagrangian point L$_1$) during quiescences ($\beta= t_{out}/t_{quiet}$), in line with the so called {\it radio-ejection} model \citep{Burderi2001a, Burderi2002a, diSalvo08, Burderi09}, in which pulsar pressure hamper accretion most of the time. Assuming a plausible mass range of the companion star (0.04--0.6 M$_{\odot}$) we isolated two extreme values of the parameter $n$: $n=1$ representing a low-mass main sequence companion and $n=-1/3$ describing a low massive evolved and fully convective companion star. Therefore, the four curves in Fig.~\ref{fig:orb_ev} refer to conservative and non-conservative scenarios each with the two mentioned values of the stellar index. In the same figure, the two horizontal lines mark the $3\sigma$ orbital period derivative confidence level interval obtained from the analysis. The vertical lines represent the minimum mass estimated from the binary mass function (assuming a 1.4  M$_{\odot}$ NS) and the companion mass for an almost face-on system (i=$10^\circ$). From Fig.~\ref{fig:orb_ev}, we deduce that a main sequence star that contracts under mass loss ($n=1$) would yield to a very small orbital period derivative for both mass loss scenarios, whilst a fully convective companion implies strong orbital expansion if the evolution is highly non-conservative as predicted e.g. by the radio-ejection scenario. Therefore, future measurements of the orbital period derivative will help constraining the secular evolution of the system.   

We investigated the NS spin frequency secular evolution by comparing the spin frequency at the end of the second outburst observed in 2008 with the value at the beginning of the latest outburst. Following \citet{Papitto11b}, the 2008 mean spin frequency was $\nu=598.89213082(2)$ Hz, with an upper limit on the spin-up derivative of $|\dot{\nu}|<4.5\times 10^{-13}$ Hz/s , estimated for an outburst duration of $\sim10$ days. Combining these information we estimated a frequency value $\nu_{2008}=598.8921308(4)$ Hz at the end of the outburst. Under the assumption that the spin value obtained from the \xmm{} observation (see Tab.~\ref{tab:solution}) is a good proxy of the NS spin of the beginning of the 2015 outburst we found $\Delta \nu = \nu_{2015}-\nu_{2008}=(1\pm 4)\times 10^{-7}$ Hz, corresponding to a $3\sigma$ confidence level interval $-11 \times 10^{-7}$ Hz $<\Delta \nu< 13\times 10^{-7}$ Hz. For a quiescence period between the two most recent outbursts of $\Delta t = 2491$ days we estimated a $3\sigma$ confidence level interval for the spin frequency derivative during quiescence $-5 \times 10^{-15}$ Hz/s $<\dot{\nu}< 6\times 10^{-15}$ Hz/s, still consistent with the spin-down derivative $\dot{\nu}_{sd}\sim -4 \times 10^{-16}$ Hz measured between the first two outbursts of the source \citep[see e.g.][]{Papitto11b, Patruno10c, Hartman2011a}.

Another interesting aspect of \igr{} is the complex behaviour of its pulse profile as function of energy. As shown in the top panel of Fig~\ref{fig:frac_amp}, the time lags resemble a piecewise function with two energy intervals where the lags increase as a function of energy (with an increment of $\sim20\mu$s in the interval $0.3-1.6$ keV and an increment of $\sim80\mu$s between $7-80$ keV), separated by the region $1.6-7$ keV where a clear decrement ($\sim170\mu$s) of the lags is observed. This result is consistent with \citet{Falanga05b}, who investigated the energy dependence of the pulse profile of \igr{} combining the observations collected by \rxte{} and \inte{}. A very similar behaviour has been observed in the AMXP IGR J17511$-$3057 \citep{Riggio11a, Falanga11}. In analogy with the lags, a complex behaviour is shown by the pulse fractional amplitude (Fig.~\ref{fig:frac_amp}, bottom panel) that increases from 6\% at 0.5 keV up to 17\% at 3 keV and then decreases down to 9\% at 10 keV. This behaviour is peculiar for the source, as generally AMXPs show also well pulsating hard components \citep[e.g.][]{Gierlinski05}. Above 10 keV, the fractional amplitude seems to increase again, even though the large uncertainties make the trend less prominent. We note that our estimates of the fractional amplitude are compatible with \citet{Falanga05b} above 10 keV. On the other hand, between 2 and 10 keV, both the amplitude and trend as a function of energy are clearly different (see their Fig.~8). This discrepancy likely reflects the large background contamination of \rxte{} due to the one degree spatial resolution. The origin of the pulse profile dependence with energy is still unclear; however, mechanisms such as strong Comptonization of the beamed radiation have been proposed to explain the hard spectrum of the pulsation, as well as the lags observed in few AMXPs \citep[e.g.][]{Patruno09B,Papitto10} including \igr{} \citep{Falanga07b}. 

\section*{Acknowledgments}
We thank the anonymous referee for helpful comments and suggestions that improved the paper.
We thank N. Schartel for providing us with the possibility to perform the ToO observation in the Director Discretionary Time, and the \xmm{} team for the technical support. We also use Director's Discretionary Time on \nustar{}, for which we thank Fiona Harrison for approving and the \nustar{} team for the technical support. We acknowledge financial contribution from the agreement ASI-INAF I/037/12/0. This work was partially supported by the Regione Autonoma della Sardegna through POR-FSE Sardegna 2007-2013, L.R. 7/2007, Progetti di Ricerca di Base e Orientata, Project N. CRP-60529. AP acknowledges support via an EU Marie Skodowska-Curie Individual fellowship under contract no. 660657-TMSP-H2020-MSCA-IF-2014, and the International Space Science Institute (ISSI) Bern, which funded and hosted the international team ``The disk-magnetosphere interaction around transitional millisecond pulsars''.

\bibliographystyle{mn2e}
\bibliography{biblio}

\label{lastpage}

\end{document}